\RequirePackage{lineno} 
\documentclass[aps,prc,twocolumn,superscriptaddress,nofootinbib]{revtex4-2}
\DeclareMathAlphabet{\mathpzc}{OT1}{pzc}{m}{it}
\usepackage{lineno}

\usepackage[T1]{fontenc}

\usepackage{graphicx}
\usepackage{amssymb}
\usepackage{amsmath}
\usepackage{amsthm}
\usepackage[percent]{overpic}
\usepackage{float}
\usepackage{array}
\usepackage{multirow}
\usepackage{epsfig}
\usepackage{color}
\usepackage{verbatim}
\usepackage{enumitem}

\usepackage{simplemargins}
\setbottommargin{2.2cm}

\usepackage[hidelinks,colorlinks=true,allcolors=blue,urlcolor=blue]{hyperref}

\newcommand{\D}{\mathrm{d}}
\newcommand{\en}{E}
\newcommand{\etof}{\mathcal{E}}

\newcommand{\x}{X}

\newcommand{\DOI}{https://doi.org/}

\begin{document}

\title{Machine learning based parametrization of the resolution function for the first experimental area (EAR1) of the n\_TOF facility at CERN}

\author{P.~\v{Z}ugec} \affiliation{Department of Physics, Faculty of Science, University of Zagreb, Zagreb, Croatia} \affiliation{European Organization for Nuclear Research (CERN), Switzerland}

\author{M.~Sabat\'{e}-Gilarte} \affiliation{STFC -- Rutherford Appleton Laboratory, Particle Physics Department, Harwell, Didcot OX11 0QX, United Kingdom} \affiliation{European Organization for Nuclear Research (CERN), Switzerland}

\author{M.~Bacak} \affiliation{TU Wien, Atominstitut, Stadionallee 2, 1020 Wien, Austria} \affiliation{European Organization for Nuclear Research (CERN), Switzerland} 

\author{V.~Vlachoudis} \affiliation{European Organization for Nuclear Research (CERN), Switzerland} 

\author{A.~Casanovas} \affiliation{Institut de T\`{e}cniques Energ\`{e}tiques (INTE)--Universitat Polit\`{e}cnica de Catalunya, Barcelona, Spain} \affiliation{Instituto de F\'{i}sica Corpuscular, CSIC--Universidad de Valencia, Valencia, Spain} \affiliation{European Organization for Nuclear Research (CERN), Switzerland} 

\author{F.~Garc\'{i}a-Infantes} \affiliation{University of Granada, Spain} \affiliation{European Organization for Nuclear Research (CERN), Switzerland} 


\begin{abstract}
This study addresses a challenge of parametrizing a resolution function of the neutron beam from the neutron time of flight facility n\_TOF at CERN. A difficulty stems from a fact that a resolution function exhibits rather strong variations in shape, over approximately 10 orders of magnitude in neutron energy. In order to avoid a need for a manual identification of the appropriate analytical forms -- hindering past attempts at its parametrization -- we take advantage of the versatile machine learning techniques. In particular, we parametrize it by training a multilayer feedforward neural network, relying on a key idea that such networks act as the universal approximators. The proof of concept is presented for a resolution function for the first experimental area of the n\_TOF facility, from the third phase of its operation. We propose an optimal network structure for a resolution function in question, which is also expected to be optimal or near-optimal for other experimental areas and for different phases of n\_TOF operation. In order to reconstruct several resolution function forms in common use from a single parametrized form, we provide a practical tool in the form of a specialized \texttt{C++} class encapsulating the computationally efficient procedures suited to the task. Specifically, the class allows an application of a user-specified temporal spread of a primary proton beam (from a neutron production process at n\_TOF) to a desired resolution function form.
\end{abstract}

\maketitle


\section{Introduction}



Neutron time of flight facility n\_TOF at CERN is a neutron production facility specializing in high-resolution measurements of the neutron induced reactions~\cite{conception,facility}. In use since 2001, it is currently in the fourth major phase of its operation~\cite{phase4,phase4_target}. Today it features three distinct experimental areas. The first and the second experimental area -- EAR1~\cite{facility} and EAR2~\cite{ear2_1,ear2_2,ear2_3} -- are well established and have long since been in use. A new NEAR~\cite{near_1,near_2,near_3} experimental area is the most recent feature, characterizing the latest n\_TOF phase.

The facility relies on a 20~GeV proton beam from the CERN Proton Synchrotron, which irradiates a massive Pb spallation target as a primary source of a neutron beam. The pulsed proton beam -- 7~ns wide (RMS), with a minimum repetition period of 1.2~s -- delivers an average of $8.5\times10^{12}$ protons per pulse. All experimental areas connect to the same spallation target. EAR1 is at a horizontal distance of approximately 185~m from the target, EAR2 is 20~m above the target, while NEAR is at the short horizontal distance of only 1.5~m from the target. The primary spallation products consist of an intense burst of $\gamma$-rays, highly energetic neutrons and the other neutral and charged particles. On their way toward EAR1 and EAR2 the charged particles are swept away by the strong electromagnets. No such magnet is used for NEAR due to its proximity to the target. Remaining ultrarelativistic spallation products reach the experimental areas as an intense burst known as the $\gamma$-flash.

Initially fast spallation neutrons are moderated by passing through a spallation target itself, through a layer of demineralized water from a cooling system and through an additional layer of borated water from a separate moderation system around the target. This yields a white neutron spectrum spanning more than 10 orders of magnitude in energy: from thermal ($\sim$10~meV) up to $\sim$1~GeV (up to the order of magnitude, depending on the experimental area~\cite{flux_ear1,flux_ear2}). The beam production, moderation and transport mechanisms are well understood~\cite{beam1,beam2}.

An inevitable by-product of the neutron production and moderation is a finite spread of neutron arrival times at the measuring station from a given experimental area, even for the neutrons of the same kinetic energy. These arrival times are measured and treated as the neutron times of flight, relative to the single initial moment of the primary proton beam hitting the spallation target. There are three major effects causing the variations in times of flight: (1)~a time width (7~ns RMS) of the primary proton beam; (2)~a distribution of neutron moderation times  inside the target-moderator assembly; (3)~a geometry of neutron transport along the beamline of finite length and breadth. This spread in neutron arrival times gives rise to a distribution known as the resolution function of the neutron beam. It causes the smearing of the experimental spectra in the cross section measurements based on the time of flight technique. As such, it must be accounted for during the analysis of the experimental time of flight data. At n\_TOF the resolution function considerations have been pursued ever since the initial conception of the facility~\cite{conception} and continue to be followed since the start of its operation~\cite{start_1,start_2} to the present day~\cite{facility,phase4_target,ear2_1,beam1,beam2}.

The only practical means of obtaining a detailed evaluation of the resolution function are the dedicated simulations of the neutron production and moderation. Due to the complexity of the target-moderator assembly at n\_TOF, these simulations are so computationally intensive that their output needs additional post-processing by the so-called optical transport code~\cite{beam1,beam2,rf_ntof}. The purpose of this code is to propagate the outgoing neutrons towards the measuring station and to refine the raw statistics from the primary simulations in a meaningful and computationally efficient way. However, the final output of this code is still subject to statistical fluctuations which are detrimental to the quality of the experimental data analysis. Furthermore, the raw numerical format of the resolution function is rather cumbersome to deal with, requiring users to implement their own interpolation and smoothing procedures. For this reason a smooth parametrization of the resolution function is highly desirable. A difficulty arises from the fact that the shape of the resolution function varies significantly over a wide energy range of the n\_TOF beam, covering more than 10 orders of magnitude in neutron energy. Attempts have been made in the past to identify the appropriate analytical form, as in~Ref.~\cite{facility}. But this form is a rather complicated function of two variables: the neutron energy and the time of flight. As such it is exceedingly difficult to identify, only to be invalidated after each modification of the neutron production system at n\_TOF; e.g. after the occasional upgrades of the spallation target, moderator assembly, beam collimation system, etc.

In this work we present an efficient and streamlined method for a parametrization of the n\_TOF resolution function by means of the machine learning techniques, together with a user-friendly interface for its evaluation. The interface consists of a dedicated \texttt{C++} class centred around the neural network implementation from a widely used programming package \textsc{ROOT}~\cite{root}. As a proof of concept, we apply the methodology to the resolution function of the first experimental area (EAR1) from the third phase (Phase-3) of the n\_TOF operation (2014--2018~\cite{phases_1,phases_2}; after a long shutdown 2019--2021, Phase-4 is in effect since 2022~\cite{phase4,phases_1}). We disclose an optimal network structure for this particular resolution function, which should also serve as the optimal or near-optimal structure for its re-parametrization after any alteration of the resolution function, or even for the parametrization of a resolution function for a different experimental area. As such, a repeated neural network training procedure requires very little user input regarding a selection of the appropriate parametrization form (realized through a selection of hyperparameters defining a neural network structure).

Section~\ref{formalism} establishes a basic formalism behind the resolution function. Section~\ref{parametrization} presents its parametrization by means of a trained neural network, together with a procedure for a numerical reconstruction of the various resolution function forms from a single parametrization. Section~\ref{conclusions} sums up the main conclusions of this work. Appendix addresses an apparent norm violation in applying the resolution function.


\section{Resolution function formalism}
\label{formalism}


A detailed resolution function formalism may be found in Ref.~\cite{rf_unfolding}. We summarize here the most important points. For readability of expressions, we will use compact notations $\en$ and $\etof$ for two different types of kinetic energy parameters. The first is a \textit{true} neutron energy~$\en$. The second is a \textit{reconstructed} neutron energy~$\etof$, calculated from a relativistic kinetic energy relation:
\begin{equation}
\etof=m c^2\left\{\left[1-\left(\frac{L}{cT}\right)^2\right]^{-1/2}-1\right\},
\label{etof}
\end{equation}
with~$m$ as the neutron mass, $c$ as the speed of light in vacuum, $L$ as a nominal neutron flight path (a length of an evacuated beamline) and, crucially, $T$ as a neutron time of flight. Let us parameterize the neutrons irradiating the sample by some kinematic parameter~$\x$. For the moment, $\x$ may be a neutron time of flight~$T$ or a reconstructed energy~$\etof$. Let $\D P_\x(\en,\x')$ be the probability for a neutron of true kinetic energy~$E$ to arrive at the sample with the specific value $\x'$ of a selected parameter~$\x$, i.e. with a value within an interval~$\D\x'$. By definition, the resolution function $R_\x(\en,\x')$ is a differential quantity:
\begin{equation}
R_\x(\en,\x')\equiv\frac{\D P_\x(\en,\x')}{\D \x'}.
\end{equation}
It is normalized such that:
\begin{equation}
\int_{-\infty}^\infty R_\x(\en,\x')\D \x'=1\quad\text{for every } \en.
\label{norm}
\end{equation}

\begin{figure}[t!]
\centering
\includegraphics[width=1\linewidth]{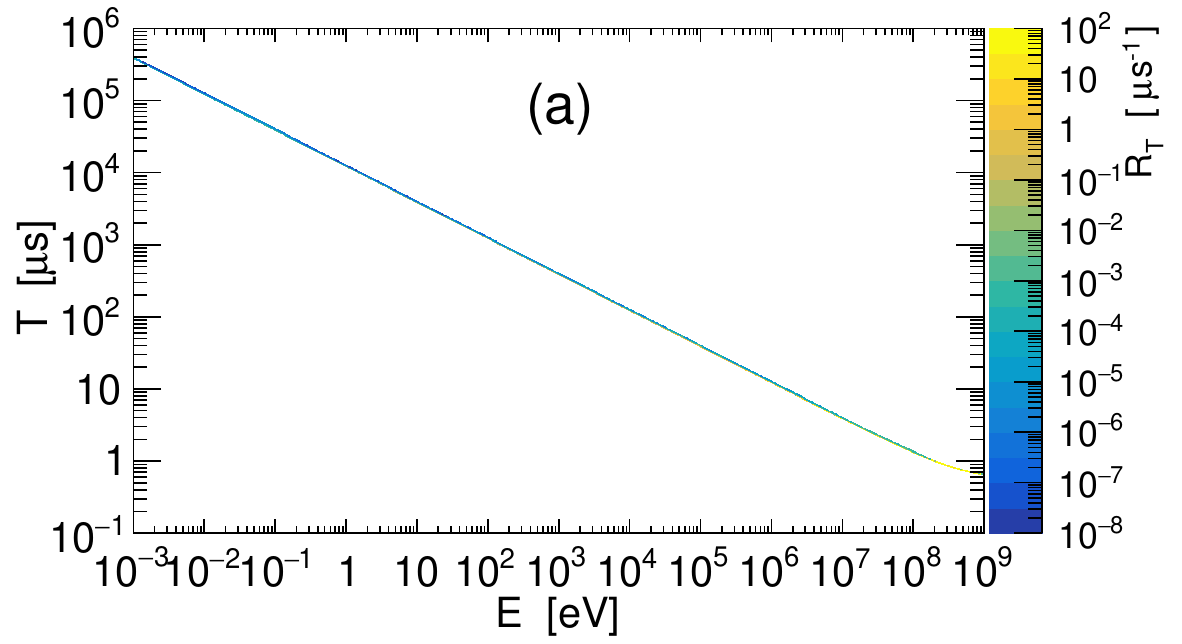}
\includegraphics[width=1\linewidth]{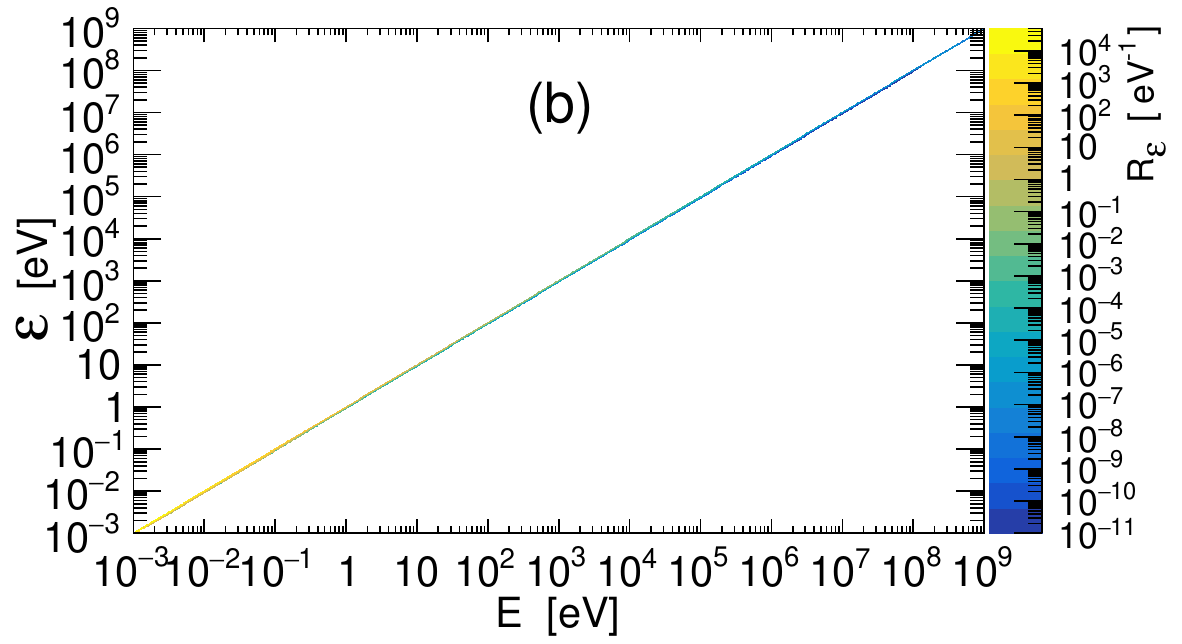}
\caption{Resolution function for the first experimental area (EAR1) of the n\_TOF facility from the third phase (Phase-3) of its operation. Top~(a): a form $R_T(\en,T')$ dependent on the neutron time of flight~$T$. Bottom~(b): a form $R_{\etof}(\en,\etof')$ dependent on the reconstructed neutron energy~$\etof$.}
\label{RF_tof_etof}
\end{figure}

The time of flight~$T$ is the most natural variable for a resolution function, due to it being directly measured in the time of flight experiments. However, both~$T$ and $\etof$ are somewhat inconvenient for a \textit{comprehensive} representation of a resolution function, for two reasons. One is that $\etof$ closely follows a true neutron energy~$E$, thus also affecting $T$ via Eq.~(\ref{etof}). As a consequence, the mean values of resolution functions $R_\etof$ and $R_T$, at given~$\en$, are closely dependent on~$E$. This is inconvenient for the neutron beams spanning multiple orders of magnitude in energy, as the relevant portion of a resolution function  is "stretched out" \textit{in both directions} throughout the parameter space. This is clearly shown in Fig.~\ref{RF_tof_etof}, which shows both forms $R_T(\en,T')$ and $R_{\etof}(\en,\etof')$ of the same resolution function.

The other reason for a cumbersome nature of $R_T$ and $R_\etof$ is the fact that the neutron time of flight directly depends on a neutron flight path~$L$, i.e. on a sample distance from a neutron source, that often changes between experiments. Thus, with every change of $L$, both $R_T$ and $R_\etof$ should be recalculated from the start. Therefore, it would be highly desirable to introduce an alternative kinematic parameter satisfying the following requirements: (1)~its span of values over the entire range of neutron energies is \textit{weakly} dependent on~$E$, being localized around some meaningful value; (2)~a resolution function in this parameter is independent of a \textit{trivial} scaling\footnote{A dependence on $L$ always has a trivial component due to the scaling of $T$ with $L$. However, a change in $L$ may also affect a resolution function in nontrivial ways due to the physical effects, such as the beam diffraction along the neutron flight path (see, for example, a short discussion around Eq.~(5) from Ref.~\cite{rf_unfolding}). While the resolution function for EAR1 shows only the trivial scaling with $L$, the one for EAR2 is strongly dependent on $L$ in a nontrivial way~\cite{rf_ntof}. This fully-trivial dependence for EAR1 is only approximate, as the resolution function is \textit{in principle} always affected by a nontrivial component. However, a very long flight path toward EAR1 (185~m) suppresses the nontrivial effects of the sample position, which was experimentally confirmed by dedicated resonance measurements.
} with~$L$, making it representative only of a neutron production process and of the nontrivial, physically meaningful effects of~$L$; (3)~preferably, a new parameter should have at least approximate physical interpretation, rather than just being an artificial mathematical transformation. This parameter has long since been identified as an \textit{effective moderation length}~$\lambda$. The idea is to separate the time of flight of monoenergetic neutrons -- that fluctuates due to the statistical nature of a neutron production and transport process -- into a contribution from a nominal flight path~$L$ and a "corrective" contribution~$\lambda$, encoding the effects of fluctuations:
\begin{equation}
v_\en T=L+\lambda,
\label{lambda}
\end{equation}
with $v_\en$ as a true neutron speed upon leaving a neutron source (a spallation target). Let us at this point introduce a following expression:
\begin{equation}
\beta_\varepsilon=\frac{\sqrt{\varepsilon(\varepsilon+2mc^2)}}{\varepsilon+mc^2},
\label{beta}
\end{equation}
since it is relevant here and will be useful in later calculations. It represents a standard relativistic factor \mbox{$\beta_\varepsilon=v_\varepsilon/c$} for neutrons of kinetic energy $\varepsilon$. As such, a true neutron speed from Eq.~(\ref{lambda}) is given by a true neutron energy as \mbox{$v_\en=c\beta_\en$}.

\begin{figure}[t!]
\centering
\includegraphics[width=1\linewidth]{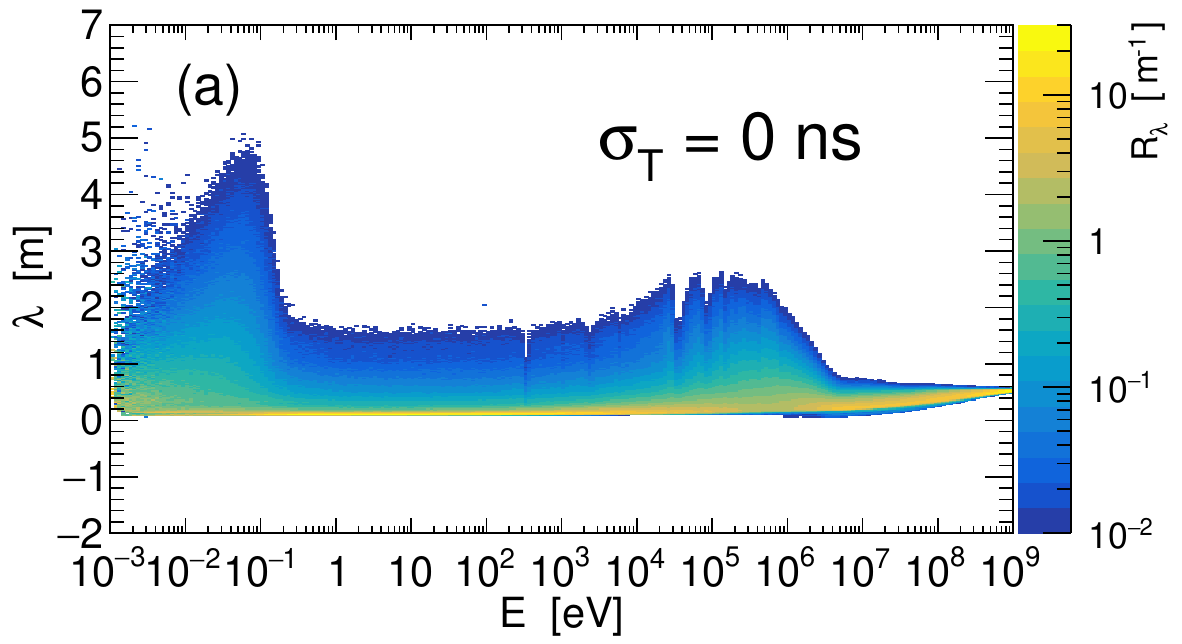}
\includegraphics[width=1\linewidth]{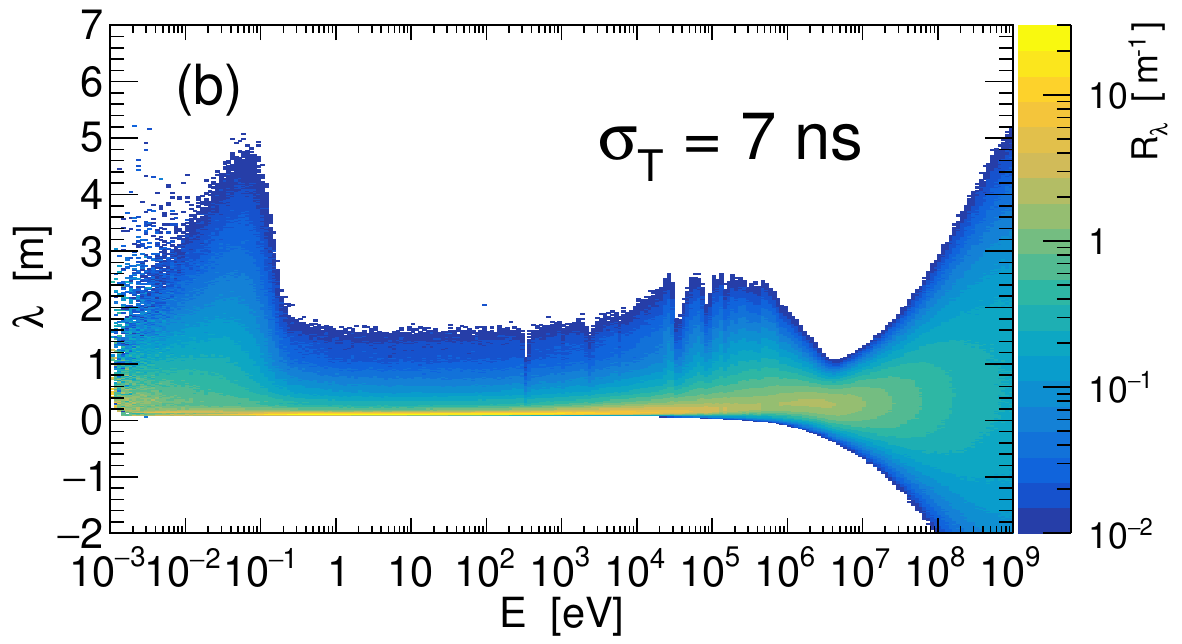}
\caption{A resolution function $R_\lambda(E,\lambda')$ dependent on the effective neutron moderation length~$\lambda$. The top form~(a) corresponds to those from Fig.~\ref{RF_tof_etof}, coming directly from the FLUKA+MNCP simulations of the neutron production and transport. In the later text -- starting with Eq.~(\ref{trans2}) -- it is denoted as~$\mathcal{R}_\lambda$. The bottom form~(b), smeared by the proton beam width \mbox{$\sigma_T=7$~ns}, corresponds to the real experimental situation and is later denoted as~$R_\lambda$.}
\label{RF_lambda}
\end{figure}

It should be noted that the effective moderation length~$\lambda$ is \textit{not} the real path length of a neutron inside a spallation target, for multiple reasons: (1)~a neutron inside a spallation target does not propagate the entire time with speed $v_\en$; (2) production of separate neutrons is initiated at different initial moments due to a finite time spread of the proton beam irradiating a spallation target, while the time of flight~$T$ for all neutrons corresponding to the same proton pulse is measured relative to a unique, fixed moment; (3)~even after leaving the spallation target, a contribution to the total~$T$ from a neutron propagation inside an evacuated beamline is not necessarily \mbox{$L/v_\en$}. This may be because a neutron is emitted at some slight angle~$\theta$ relative to the beamline axis, making a real flight path \mbox{$L/\cos\theta$}. It may also scatter off the beamline walls, which increases its flight path \textit{and} alters its speed.

Figure~\ref{RF_lambda} shows a resolution function $R_\lambda(E,\lambda')$ dependent on the effective moderation length. The top form (\mbox{$\sigma_T=0$}) does not take into account the time width $\sigma_T$ of the primary proton beam (explained later) and perfectly corresponds to the earlier forms from Fig.~\ref{RF_tof_etof}. For convenience we immediately show a resolution function smeared by the proton beam RMS of \mbox{$\sigma_T=7$~ns}, which will soon be elaborated.



We now have a set of three kinematic parameters to be found in common use: \mbox{$\x\in\{T,\etof,\lambda\}$}. The transformation of a resolution function between these parameters follows from a conservation of probability:
\begin{equation}
R_T(\en,T')\;|\D T'|=R_{\etof}(\en,\etof')\;|\D \etof'|=R_\lambda(\en,\lambda')\;|\D \lambda'|.
\label{trans1}
\end{equation}
It is shown in Ref.~\cite{rf_unfolding} that the resolution function transforms a \textit{differential spectrum} of counts $N_\en(E')$ dependent on a true neutron energy into a differential spectrum of counts $N_\x(\x')$ dependent on a selected kinematic parameter as:
\begin{equation}
\frac{\D N_\x(\x')}{\D\x'}=\int_0^\infty \frac{\D N_\en(\en')}{\D\en'} R_\x(\en',\x')\D\en'.
\label{counts_trans}
\end{equation}

\section{Resolution function parametrization by means of machine learning}
\label{parametrization}


\subsection{Resolution function fitting}

It was shown in Ref.~\cite{rf_ntof} that the raw resolution function (obtained by the optical code) can not be used for a reliable resonance analysis. The reason is that the residual statistical fluctuations from the computationally intensive FLUKA+MNCP simulations of the neutron production and transport through the spallation target are not negligible relative to the fluctuations in the experimental data. Thus, using the raw resolution function in the analysis of the experimental data artificially and unnecessarily increases the involved statistical uncertainties. (Examples of smearing the initially smooth spectra by the raw resolution function may be found in later Figs.~\ref{fig_resonances} and~\ref{fig_unit_spectrum}.) Clearly, a smoothed form of the resolution function is necessary, so as to avoid this adverse effect. It is also highly desirable that the smoothed form be efficiently parameterized, i.e. that it be more compact than just a densely interpolated resolution function matrix filled with the values of a smoothed function (a matrix such as those from Figs.~\ref{RF_tof_etof} and~\ref{RF_lambda}).

One way of proceeding would be to identify an analytical parametrization of the entire resolution function matrix. An example of such parametrization for a resolution function of EAR1 from Phase-1 of the n\_TOF operation may be found in Ref.~\cite{facility}. However, such analytical form is difficult to identify and may no longer be appropriate when the alterations are introduced to the neutron production process. For example, the replacement of a spallation target between Phase-1 and Phase-2 of the n\_TOF operation~\cite{phase4_target} notably affected the shape of a resolution function, rendering previous parametrization invalid. Furthermore, a resolution function for EAR2 differs from the one for EAR1 and requires its own dedicated parametrization. In order to avoid a tedious manual identification of new analytical forms, we take advantage of the machine learning techniques, in particular of the deep feedforward neural networks. The idea stems from a fact that the multilayer feedforward neural networks act as the \textit{universal approximators}, capable of approximating any sufficiently well behaved function to any desired degree of accuracy~\cite{approximators1,approximators2}. In other words, such networks can be thought of as "black box" fitting functions, capable of modeling any function of practical importance. The application of neural networks to this task is a part of ongoing efforts to introduce the machine learning techniques into a widespread practice at n\_TOF~\cite{neural_2021,neural_2022,neural_2023}. The possibility of applying the convolutional neural networks in unfolding the effects of the resolution function is also being investigated, with very promising results on the horizon~\cite{rf_neural}. 


We demonstrate the proof of concept by fitting a resolution function of EAR1, from Phase-3 of the n\_TOF operation. To this end use the neural network training capabilities of \texttt{TMultiLayerPerceptron} class~\cite{root_perceptron} from \textsc{root}. Using \textsc{root} allows for a seamless integration of the end result (a trained neural network) within a vast majority of the data analysis codes from n\_TOF. We provide a basic example of the code usage among the openly available data files~\cite{data_bank}.

Our goal is to fit a single form of a resolution function (either $R_T$, $R_\etof$ or $R_\lambda$) and to reconstruct all other forms from this single fit by applying the appropriate transformations. This will ensure a perfect consistency between all forms of a resolution function, which would not necessarily be satisfied by fitting each form separately. Due to the described advantages of the $R_\lambda$ representation (a uniformity of relevant $\lambda$~values and insensitivity to a nominal flight path~$L$), it is an obvious choice for fitting.

There is another consideration to be taken into account, that will allow for a greater flexibility in reconstructing particular forms of a resolution function. Resolution functions of the n\_TOF facility (for different experimental areas) are affected by two \textit{separable} contributions: (1)~a neutron production and transport process inside a spallation target, as well as a neutron transport outside of it; and (2)~a time distribution (a finite time width) of the primary proton beam from the CERN Proton Synchrotron irradiating the spallation target. The proton beam time distribution is Gaussian in shape, with a standard deviation of \mbox{$\sigma_T=7$~ns}. Let $\mathcal{R}_T$ designate a resolution function in time of flight, \textit{without} the effects of the proton beam width (as if $\sigma_T=0$). $\mathcal{R}_T$ is easily extracted from the raw results of the FLUKA+MNCP simulations processed by an optical transport code. A resolution function $R_T$ \textit{affected} by the proton beam width is then obtained by a simple convolution with a temporal proton beam profile (a normalized Gaussian):
\begin{equation}
R_T(E,T')=\tfrac{1}{\sigma_T\sqrt{2\pi}}\int_{-\infty}^\infty \mathcal{R}_T(E,\tau) \exp\left[-\tfrac{(T'-\tau)^2}{2\sigma_T^2}\right]\D \tau.
\label{convolution_tof}
\end{equation}

On account of a linear relationship between $T$ and $\lambda$ from Eq.~(\ref{lambda}), a representation $R_\lambda$ of a resolution function affected by a proton beam width may still be expressed as a convolution of a resolution function $\mathcal{R}_\lambda$ with an instantaneous proton beam, and a $\lambda$-equivalent of a temporal beam profile. From Eq.~(\ref{lambda}) it follows that the $\lambda$-width of this profile is dependent on a true neutron energy as \mbox{$\sigma_\lambda(E)=v_\en \sigma_T$}. Using Eq.~(\ref{beta}) this dependence may be expressed as:
\begin{equation}
\sigma_\lambda(E)=c\beta_E \sigma_T=\frac{\sqrt{E(E+2mc^2)}}{E+mc^2}\,c\sigma_T.
\label{sigma_lambda}
\end{equation}
Thus a required convolution takes a form:
\begin{equation}
R_\lambda(E,\lambda')=\tfrac{1}{\sigma_\lambda(E)\sqrt{2\pi}}\int_{-\infty}^\infty \mathcal{R}_\lambda(E,\Lambda) \exp\left[-\tfrac{(\lambda'-\Lambda)^2}{2\sigma_\lambda^2(E)}\right]\D \Lambda.
\label{trans2}
\end{equation}

Equations~(\ref{trans1}) and~(\ref{trans2}) constitute all the transformation rules required for reconstructing any desired form of a resolution function from a proton-beam-width-free form~$\mathcal{R}_\lambda$. Therefore, we apply a neural network fitting precisely to this, most convenient representation~$\mathcal{R}_\lambda$. The arguments $E$ and $\lambda'$ of a two-dimensional resolution function $\mathcal{R}_\lambda(E,\lambda')$ play a role of the network inputs. A single scalar-function value for each pair of arguments plays a role of a single network output. However, since the true neutron energies~$E$ and the resolution function values~$\mathcal{R}_\lambda$ span multiple orders of magnitude, for reasons of numerical stability we take the logarithms of these quantities for the input and output neutrons.

\begin{figure}[t!]
\centering
\includegraphics[width=1\linewidth]{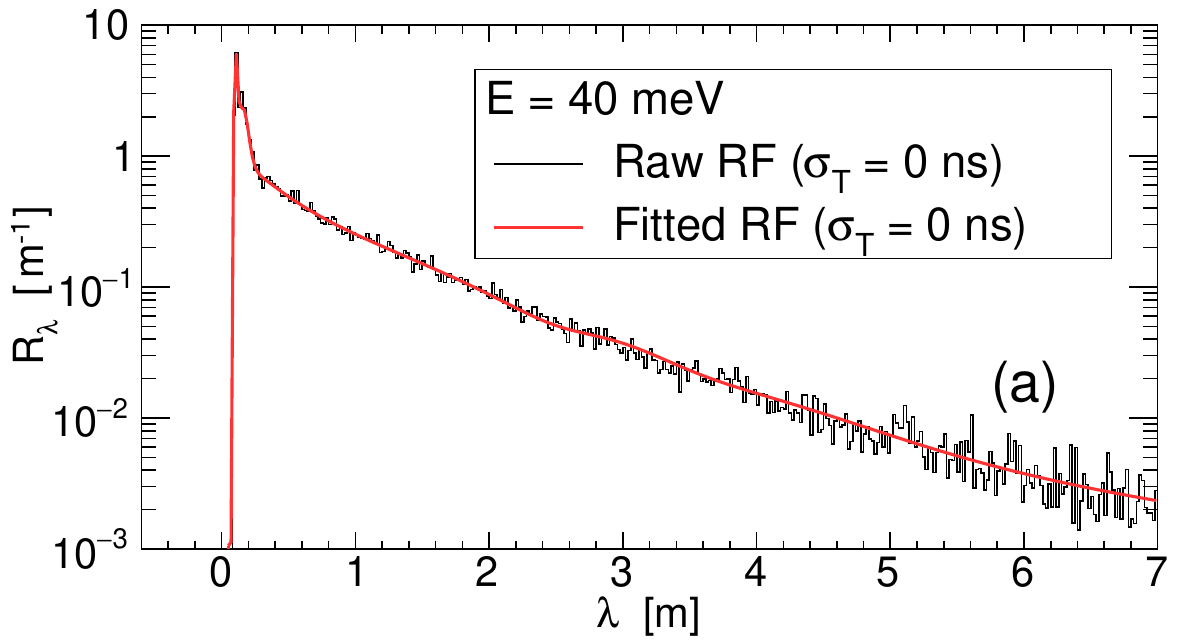}
\includegraphics[width=1\linewidth]{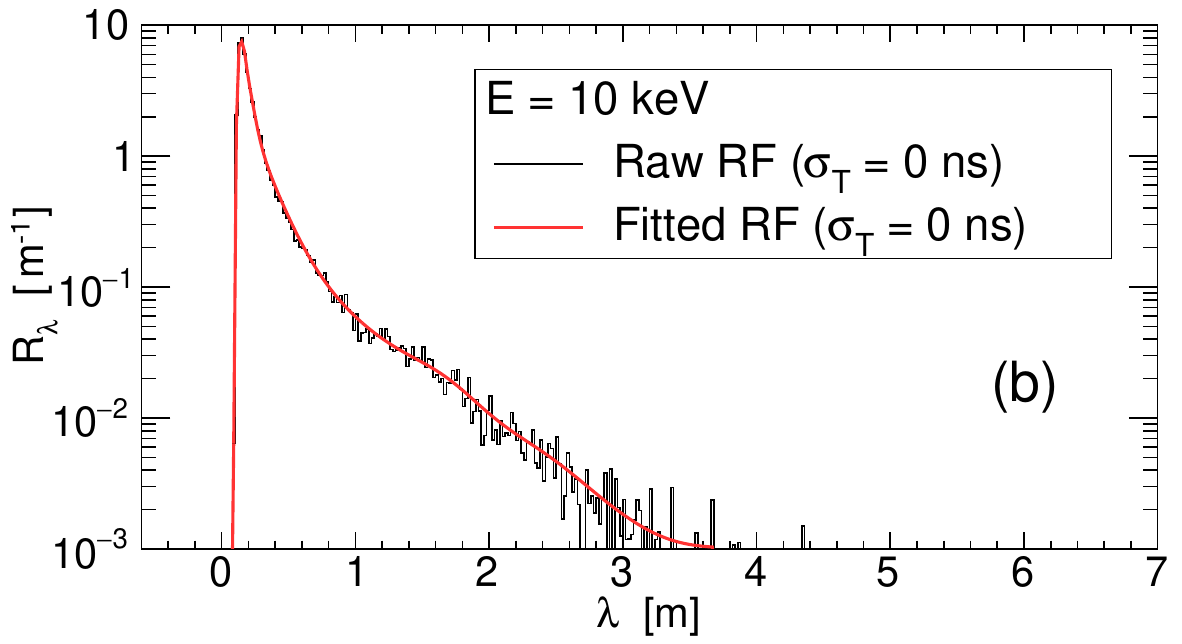}
\includegraphics[width=1\linewidth]{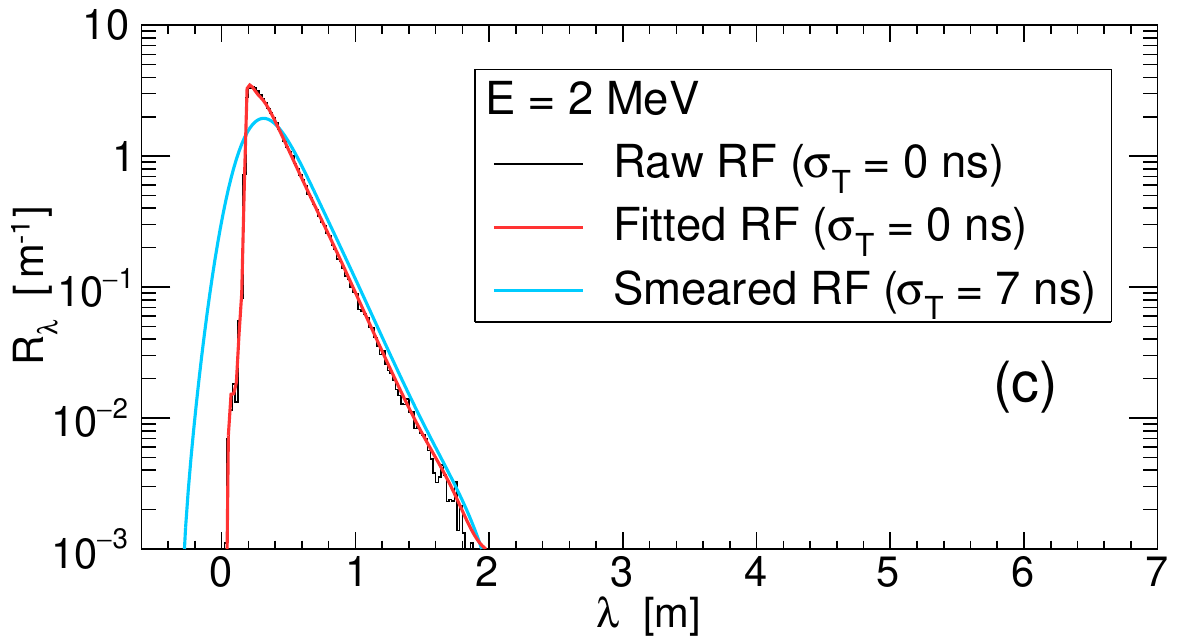}
\caption{Raw resolution function fitted by a single neural network, at neutron energies of (a)~40~meV, (b)~10~keV and (c)~2~MeV. A raw and fitted function correspond to an \textit{instantaneous} proton beam irradiating a spallation target. A beam width of {$7$~ns} does not have a notable effect below 10~keV.
}
\label{fig_slices}
\end{figure}

We describe here a neural network training and optimization procedure. First, the 450$\times$60 numerical resolution function matrix was constructed, spanning 450 uniformly distributed $\lambda$ values between $-$2~m and 7~m (2~cm steps), together with 60 isolethargically distributed $\log_{10}E$ values between 10$^{-3}$~eV and 10$^9$~eV (5 points per decade), creating a dataset of \mbox{27\,000} points. It should be noted that the trained neural network \textit{is not be used for any kind of extrapolation outside of his range}. It serves only as a smooth and compact parametrization throughout the physically meaningful range, to be used only for the resolution function reconstruction within a range entirely covered by the numerical matrix being fitted. Therefore, no cross-validation was required during a training procedure. In other words, all \mbox{27\,000} entries could be used as training counts, without needing to reserve any of them for separate testing. Thus the quality of the trained network could be entirely assessed against the training data. We used a few simple and effective steps: we observed the saturation rate of a loss function during the training, and visually compared the agreement between the final fit and the raw data (see Fig.~\ref{fig_slices}, soon to be discussed). We also closely monitored a distribution of fitting residuals.

\begin{figure}[t!]
\centering
\includegraphics[width=0.75\linewidth]{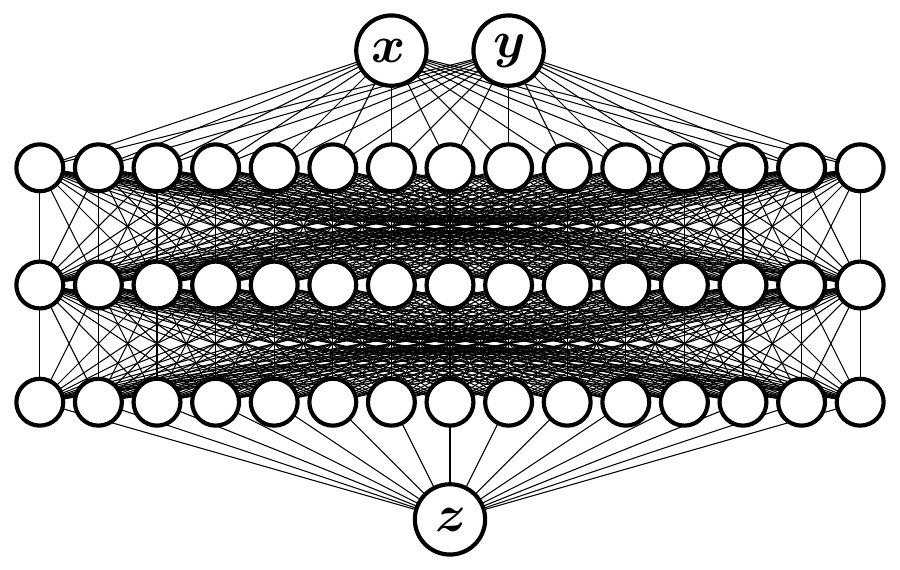}
\vspace*{-2mm}
\caption{Neural network structure used for modeling a resolution function of EAR1 from Phase-3 of n\_TOF operation. Each fully connected hidden layer consists of 15 neurons. Inputs correspond to \mbox{$x=\lambda'/\text{(1 cm)}$} and \mbox{$y=\log_{10}[E/\text{(1 eV)}]$}. A single output is \mbox{$z=\log_{10}[\mathcal{R}_\lambda/(\text{1 cm}^{-1})]$}.} 
\vspace*{-2mm}
\label{fig_nn_structure}
\end{figure}

We tested several training methods available in the \texttt{TMultiLayerPerceptron} class, using a few preliminary network structures. We did not observe any significant improvement -- either in the computational efficiency or in the quality of the final results -- over the default Broyden-Fletcher-Goldfarb-Shanno (BFGS) method with the default hyperparameter values~\cite{root_perceptron}. Hence, we opted for the default training parameters, employing the sigmoid activation function. In \texttt{TMultiLayerPerceptron} the binary cross-entropy is the only loss function associated with the sigmoid activation function. By testing a range of network structures we identified an optimal structure consisting of three hidden layers, each composed of 15 neurons. The optimal structure was easily identified. Less than three hidden layers do not seem to have enough flexibility to recover the quick variations in the resolution function, even with highly increased number of neurons. Aside from the visually obvious overfitting, more than three hidden layers do not bring any further improvement in the resolution function reconstruction. Once the optimal three-layer structure was identified, the number of neurons was varied in steps of 5, in different combinations throughout the layers (e.g. \mbox{15-10-20}). By searching for the simplest structure providing a satisfactory resolution function reconstruction -- without further improvement with increasing number of neutrons -- we quickly converged upon the optimal \mbox{15-15-15} structure, which is shown in Fig.~\ref{fig_nn_structure}. The network inputs are \mbox{$x=\lambda'/\text{(1 cm)}$} and \mbox{$y=\log_{10}[E/\text{(1 eV)}]$}, with a single output \mbox{$z=\log_{10}[\mathcal{R}_\lambda/(\text{1 cm}^{-1})]$}. In general, the final state of a trained neural network depends on a particular training run due to a random initialization of its weights and biases. With the optimal network structure and after a sufficient number of training epoch these variations are essentially negligible. For the final training we used 10$^4$ training epochs. A single-threaded training on AMD Ryzen~7 7735HS (3.2 GHz) CPU under Linux Mint 21.2 Cinnamon takes 1~minute per 100~epochs, making a total of 100~minutes for 10$^4$~epochs.

Figure~\ref{fig_slices} compares a raw resolution function with a fitted one, for several values of a true neutron energy~$\en$. The effect of a proton beam width is negligible below $\sigma_\lambda=1$~cm. As per Eq.~(\ref{sigma_lambda}), a time width of \mbox{$\sigma_T=7$~ns} implies an equivalent energy limit of \mbox{$\en=10$~keV} (corresponding, for EAR1 flight path of \mbox{$L=180$~m}, to the times of flight below 0.13~ms). Hence, at lower neutron energies the proton beam width of 7~ns does not have any significant effect upon the resolution function. For this reason a resolution function smeared by means of a numerical convolution from Eq.~(\ref{trans2}) is shown only at energies higher than 10~keV.

\subsection{Numerical resolution function reconstruction}

In order to reconstruct the resolution function forms $R_T$ and $R_\etof$, smeared by a proton beam width, one relies on the transformation rules from Eq.~(\ref{trans1}). In general case, a smeared resolution function $R_\lambda$ should \textit{first} be obtained from an unsmeared fit $\mathcal{R}_\lambda$ -- by means of Eq.~(\ref{trans2}) -- and only then should a required transformation from Eq.~(\ref{trans1}) be applied. The reason is this. Combining Eqs.~(\ref{etof}) and~(\ref{lambda}) yields a \textit{nonlinear} relation between $\etof$ and $\lambda$:
\begin{equation}
\etof=m c^2\left\{\left[1-\left(\frac{v_E}{c}\frac{L}{L+\lambda}\right)^2\right]^{-1/2}-1\right\}.
\label{etof_lambda}
\end{equation}
Due to this nonlinearity the transformation between an unsmeared resolution function $\mathcal{R}_\etof$ and a smeared~$R_\etof$ can no longer be expressed as a formal convolution, which bears upon the questions of computational complexity in applying a smearing transformation. We address these questions in Section~\ref{rf_class}.

Due to a simple relation between $\lambda$ and $T$ from Eq.~(\ref{lambda}), a transformation rule for $R_T$ from Eq.~(\ref{trans1}) involves a very simple derivative \mbox{$\D\lambda/\D T=v_E$}, yielding:
\begin{equation}
R_T(\en,T')=c\beta_\en \times R_\lambda(\en, \, c\beta_\en T'-L),
\label{trans_tof}
\end{equation}
with $\beta_E$ being defined by Eq.~(\ref{beta}). A transformation for $R_\etof$ is slightly more complex. By first obtaining \mbox{$\lambda(\etof)$} and \mbox{$\D\lambda/\D\etof$} from Eq.~(\ref{etof_lambda}), one arrives at:
\begin{equation}
R_\etof(\en,\etof')=\frac{Lm^2c^4}{(\etof' + mc^2)^3}\frac{\beta_\en}{\beta_{\etof'}^3} \times R_\lambda\left[E, \, L\left(\frac{\beta_\en}{\beta_{\etof'}}-1\right) \right],
\label{trans_etof}
\end{equation}
where \mbox{$\lambda(\etof)=L(\beta_\en/\beta_{\etof}-1)$}. By construction, these transformations preserve a norm of a resolution function.


\begin{figure}[b!]
\centering
\includegraphics[width=1\linewidth]{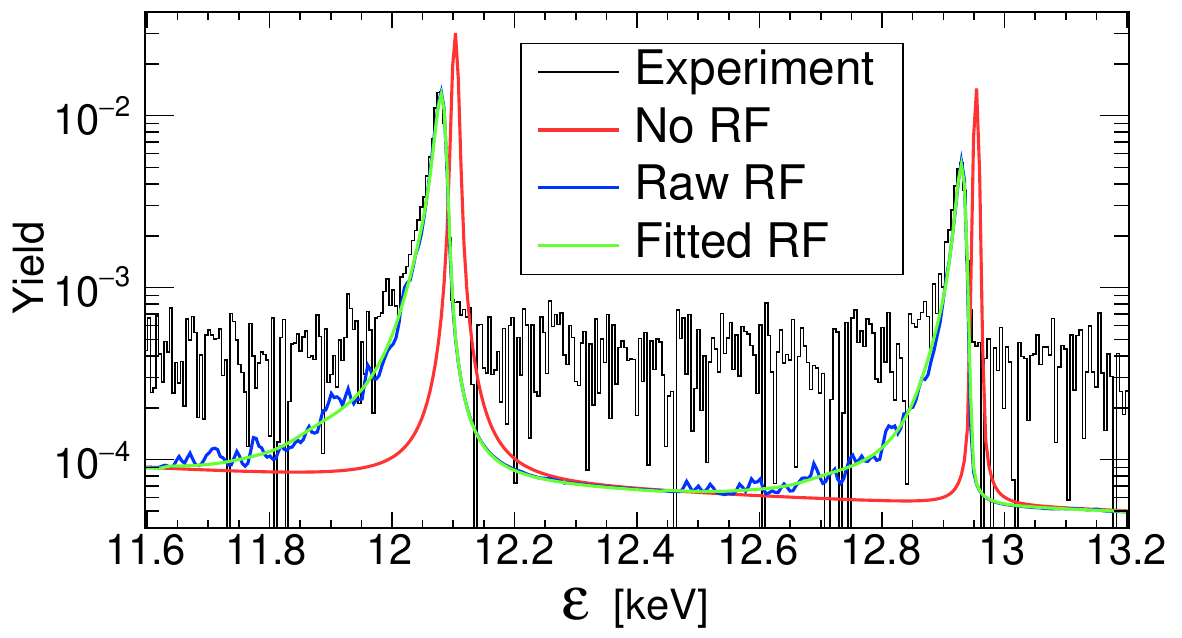}
\caption{Selected resonances from the n\_TOF measurement of the $^{53}$Cr($n,\gamma$) reaction, compared to the reaction yields based on ENDF/B-VIII.0 data, with of without the resolution function having been applied.}
\label{fig_resonances}
\end{figure}

We use a transformed resolution function from Eq.~(\ref{trans_etof}) to demonstrate an effect upon and an agreement with the n\_TOF experimental data. Figure~\ref{fig_resonances} shows two selected resonances from a recent measurement of the $^{53}$Cr($n,\gamma$) reaction in EAR1~\cite{cr53}. Though the measurement was performed during Phase-4 of the n\_TOF operation, for presentation purposes we use here a resolution function from Phase-3 as a first approximation of the resolution function from Phase-4. The plot shows a resolution-function-free reaction yield -- manually constructed by appropriately scaling a neutron capture cross section from ENDF/B-VIII.0 database~\cite{endf8} -- together with two resolution resolution-function-smeared yields, comparing them to the experimental data. (The preliminary experimental data are shown, as their analysis is not yet complete. The region between the resonances is still affected by the residual background contributions, not all of which have yet been subtracted. The relative background contribution inside the resonances is negligible for visual purposes, so that a meaningful visual comparison with the ENDF resonances can still be made.) Smeared yields have been obtained by applying either the raw, unparameterized resolution function or the one fitted by the neural network. Reaction yields ($Y$) transform precisely as the differential spectra from Eq.~(\ref{counts_trans}) and have been obtained by a transformation:
\begin{equation}
Y_\etof(\etof')=\int_0^\infty Y_\en(\en') R_\etof(\en',\etof')\D\en',
\label{yield}
\end{equation}
wherein $Y_E$ is a yield constructed from ENDF/B-VIII.0 data. Resolution function~$R_\etof$ was calculated according to Eq.~(\ref{trans_etof}), starting either from the raw or the fitted resolution function~$R_\lambda$. The difference between two smeared yields illustrates our initial claim that the raw resolution function should not be used in the accurate data analysis, as it unnecessarily introduces artificial fluctuations into otherwise smooth data, or enhances the existing ones.

\begin{figure}[t!]
\centering
\includegraphics[width=1\linewidth]{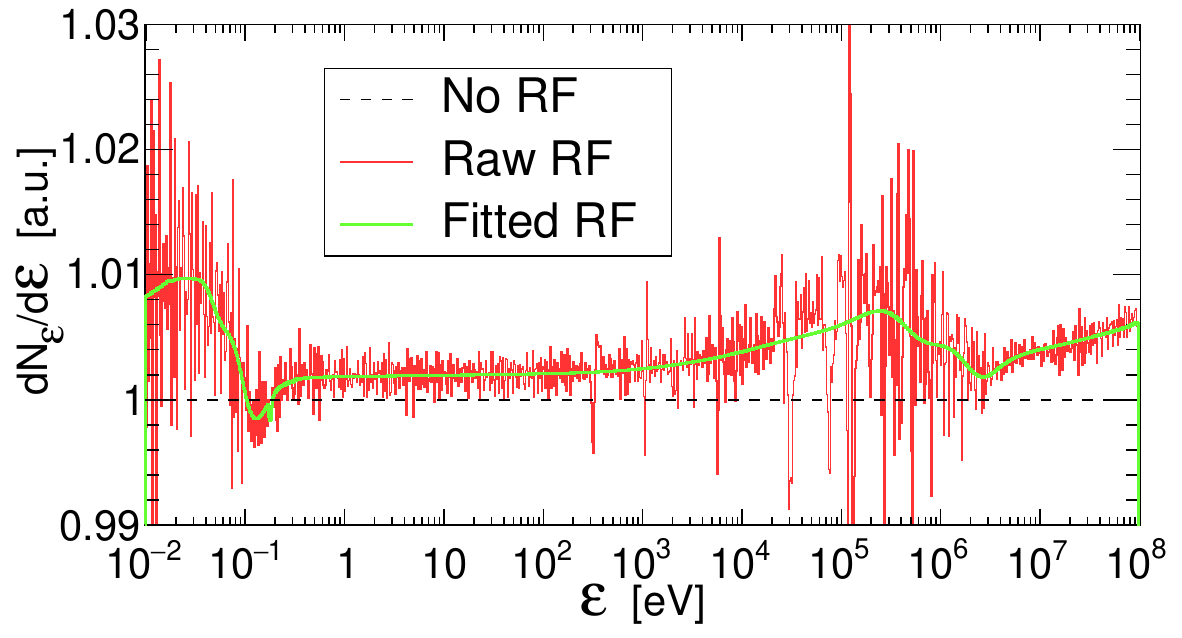}
\caption{Effect of applying either the raw or the fitted resolution function to a unit spectrum spanning 10 orders of magnitude in neutron energy.  The data are shown in 100 bins per decade. A level of fluctuations introduced by the raw resolution function increases with denser binning.}
\label{fig_unit_spectrum}
\end{figure}

A simple and comprehensive insight into effects of applying either the raw or the smoothed resolution function may be obtained by applying them to a constant spectrum of unit height. Figure~\ref{fig_unit_spectrum} shows these results for the $\etof$-spectrum spanning 10 orders of magnitude in neutron energy, completely analogously to a resonant yield from Fig.~\ref{fig_resonances}. A detrimental effect of applying the raw resolution function is immediately evident, as opposed to a smooth end-result from the fitted resolution function. In that, Figure~\ref{fig_unit_spectrum} shows the result of applying a resolution function unsmeared by a proton beam width (as if \mbox{$\sigma_T=0$}). The reason is that smearing the raw resolution function smooths a high energy part, thus obscuring the statistical fluctuations inherent in the simulated data, while these fluctuations are precisely what we are trying to exhibit here. Other than smoothing a high energy part of the spectrum obtained with the raw resolution function, a realistic value of \mbox{$\sigma_T=7$~ns} does not affect the shape of the folded spectra in a visible way (as they are displayed in Fig.~\ref{fig_unit_spectrum}). It should be taken into account that the data from Fig.~\ref{fig_unit_spectrum} are displayed in 100 bins per decade. With finer binning the fluctuations from the raw resolution function become even more pronounced. We note in passing that the n\_TOF data are often analyzed in thousands of bins per decade (see, for example, Refs.\cite{bpd_5000_1,bpd_5000_2} for 5000 bins per decade applied to the neutron capture data or Ref.~\cite{bpd_2000} for 2000 bins per decade applied to the fission data). While the resonance plots like those from Fig.~\ref{fig_resonances} show how the underlying data are \textit{locally} deformed by a resolution function, a plot from Fig.~\ref{fig_unit_spectrum} clearly shows that the \textit{global} data trend may also be affected. 

The resolution function affected spectra from Fig.~\ref{fig_unit_spectrum} seem to be systematically higher than unity. This might suggest that the application of a resolution function violates a norm preservation, i.e. that it violates a conservation of the total number of underlying counts. This violation is only apparent and is addressed in Appendix.



\subsection{Resolution function class}
\label{rf_class}


In order to facilitate the use of a newly fitted resolution function at n\_TOF, we have written a self-contained \texttt{C++} class~\cite{rf_guide}, serving as a simple interface for the evaluation of any desired resolution function form ($R_\lambda$, $R_T$ or $R_\etof$), starting from a fitted proton-beam-width-free form~$\mathcal{R}_\lambda^*$ (not necessarily normalized, as denoted by asterisk). The class serves the following functions:
\begin{itemize}[noitemsep,topsep=1pt]
\item[(1)] initialization of a trained neural network ($\to \mathcal{R}_\lambda^*$);
\item[(2)] numerical normalization ($\to \mathcal{R}_\lambda$);
\item[(3)] smearing due to a proton beam width ($\to R_\lambda$);
\item[(4)] transformation to alternative forms ($\to R_T$ or $R_\etof$);
\item[(5)] computationally efficient implementation of the above operations.
\end{itemize}
The class houses the parameters of a trained neural network and initializes a \texttt{TMultiLayerPerceptron} object, allowing for a direct evaluation of a fitted, proton-beam-width-free (unnormalized) resolution function~$\mathcal{R}_\lambda^*$.

In evaluating any form of a resolution function, the first operation to be performed is a numerical normalization such that $\int_{-\infty}^\infty \mathcal{R}_\lambda(\en,\lambda')\D\lambda'=1$, for any required value of~$\en$. The reason for performing a normalization \textit{at all} is the fact that a fitting function without \textit{a priori} imposed norm does not necessarily preserve a norm of the fitted data. Therefore, even if already-normalized data were fitted, the end-result needs to be \textit{a posteriori} normalized. This also allows for the unnormalized raw data to be fitted, which may improve the quality of the fit, since the raw data may be unnormalized not only by a constant factor, but by an arbitrary $\en$-dependent function, affecting a global twodimensional trend that the trained neural network needs to reproduce. The reason for a numerical normalization to be performed \textit{first} is a computational efficiency, owing to the fact that both the convolutional smearing from Eq.~(\ref{trans2}) and the kinematic parameter transformations from Eqs.~(\ref{trans_tof}) and~(\ref{trans_etof}) preserve the norm. To clarify the point, let $\mathbb{T}_\x$ denote a total transformation (a composition of smearing and kinematic-parameter-transformation operations) acting on unnormalized $\mathcal{R}_\lambda^*$ so as to produce an unnormalized, smeared and transformed resolution function $R_\x^*$, in a sense: \mbox{$R_\x^*=\mathbb{T}_\x\{\mathcal{R}_\lambda^*\}$}. Due to a norm preserving property of all operations from $\mathbb{T}_\x$, a \textit{normalized} resolution function may obtained in two ways:
\begin{align}
\begin{split}
R_\x&(\en,\x')=\\
&=\mathbb{T}_\x \left\{\tfrac{\mathcal{R}_\lambda^*[E,\lambda'(\x')]}{\sum_\Lambda \mathcal{R}_\lambda^*(E,\Lambda)\Delta\Lambda}\right\} 
=\tfrac{\mathbb{T}_\x\{\mathcal{R}_\lambda^*[E,\lambda'(\x')]\}}{\sum_x \mathbb{T}_\x\{\mathcal{R}_\lambda^*[E,\lambda(x)]\}\Delta x},
\end{split}
\label{comute}
\end{align}
where the sum from either denominator represents a discrete numerical integration. When the normalization is performed first, as in the first expression, the computationally expensive operations from $\mathbb{T}_\x$ need to be performed only once in order to obtain a single point from~$R_\x$. If, on the other hand, the normalization was performed last (the second expression), then the same computationally expensive operations need to be performed unnecessarily many times during the norm calculation by means of a discrete summation. Furthermore, evaluating a trained network with many neural links (Fig.~\ref{fig_nn_structure}) is also moderately expensive. For that reason we trigger a fresh normalization only when a new value of $\en$ is registered, different from a previous one requested by user. This avoids the unnecessary repetitions of the same summation from Eq.~(\ref{comute}).



For a user-supplied value of a temporal proton beam width~$\sigma_T$, a normalized resolution function $R_\lambda$ is smeared by performing a discrete version of a convolution from Eq.~(\ref{trans2}). This is the most significant bottle-neck of the numerical calculations, since a naive convolution algorithm is computationally expensive. When users request single points of a resolution function (one point at a time), little can be done to speed up a convolution algorithm itself. In this case our class offers a possibility of a bilinear interpolation, soon to be described. However, significant algorithmic improvements are available when a convolution needs to be calculated (i.e. smearing needs to be performed) over a grid of uniformly spaced $\lambda$-points. While a naive convolution algorithm is of excessive \mbox{$\mathcal{O}(N^2)$} computational complexity -- $N$~being a number of required points -- the famous Fast Fourier Transform algorithm manages a job in only \mbox{$\mathcal{O}(N\log_2 N)$} steps~\cite{fft}. In this case we employ an efficient Fast Fourier Transform implementation from Ref.~\cite{numc}, keeping the overall computational workload at a manageable level.

Having thus obtained a smeared~$R_\lambda$, only a kinematic parameter transformation from Eq.~(\ref{trans_tof}) or~(\ref{trans_etof}) remains to be performed whenever the resolution function forms $R_T$ or $R_\etof$ are requested. If the evaluation of $R_T$ over a grid of uniformly spaced $T$-points is needed, the computational advantages of the Fast Fourier Transform may again be relied on. This is because the smeared $R_T$ may still be expressed as a formal convolution; in fact, it is \textit{the} original convolution from Eq.~(\ref{convolution_tof}). At the same time, due to a linearity between~$\lambda$ and~$T$ from Eq.~(\ref{lambda}), a uniform grid of $T$-points corresponds to a uniform grid of $\lambda$-points, allowing for the Fast Fourier Transform to be applied. However, due to the n\_TOF beam spanning more  than 10 orders of magnitude in neutron energy, the measured time of flight spectra are spread over the multiple orders of magnitude. For this reason the isolethargic spacing of $T$-points is commonly used, corresponding to an isolethargic spacing of $\lambda$-points (at a given~$E$), for which a regular Fast Fourier Transform algorithm can no longer be used. On the other hand, whether a uniform of isolethargic spacing of $\etof$-points is used (as is common practice), \textit{neither} corresponds to a uniform spacing of $\lambda$-points, due to a nonlinear relationship from Eq.~(\ref{etof_lambda}). In order to extend a computational efficiency to all cases, our class allows users to activate the interpolation mode, allowing any form of the resolution function to be evaluated by means of a \textit{bilinear interpolation} between pre-calculated resolution function points. In that, a smeared and normalized resolution function~$R_\lambda$ is evaluated on a dense \mbox{$(\en,\lambda)$} grid, allowing all the normalization and smearing operations to be performed in a single go, without later repetitions. This grid is isolethargically spaced over~$\en$ and, more importantly, uniformly spaced over~$\lambda$, so that the Fast Fourier Transform can be taken full advantage of during a smearing stage. The user-requested values of a resolution function are calculated by first obtaining a required \mbox{$R_\lambda(E,\lambda')$} point by a bilinear interpolation between 4 closest points from a pre-calculated grid (a simple bilinear interpolation algorithm may also be found in Ref.~\cite{numc}). If necessary, a kinematic parameter transformation from Eq.~(\ref{trans_tof}) or~(\ref{trans_tof}) is then performed. For a sufficiently dense grid of pre-evaluated points it makes little numerical difference whether the kinematic parameter transformation is performed before or after the interpolation. However, prior to interpolation it has to be executed at 4 grid-points; after the interpolation it needs to be applied only once. We have selected a computationally efficient procedure. 

\section{Conclusions}
\label{conclusions}

We have provided an efficient way of parametrizing the resolution function of the neutron beam from the n\_TOF facility, thus solving a long-standing problem of facilitating its use among the users requiring the resolution function in their data analyses. The method takes advantage of the machine learning techniques. Specifically, the resolution function is fitted by training a multilayer feedforward neural network, due to the fact that such networks act as the universal approximators. We have applied the method to the resolution function for the first experimental area of the n\_TOF facility, from the third phase of its operation. In order to re-parametrize the resolution function after any alteration -- or to parametrize a resolution function for a different experimental area -- one only needs to retrain a neural network by using the readily available streamlined procedures. In this work we have used the neural network training capabilities of the \texttt{TMultiLayerPerceptron} class from a \texttt{C++} based programming package \textsc{root}. 

\pagebreak

We have parametrized a single most appropriate form of the resolution function, dependent on the so-called effective neutron-moderation length, and unaffected by the temporal spread of the primary proton beam from a neutron production process at n\_TOF. In order to efficiently reconstruct several other resolution function forms in common use -- those dependent on the neutron time of flight or the so-called reconstructed neutron energy -- and to apply the effects of the proton beam width, we have supplied a standalone \texttt{C++} class specializing in those tasks. The class is immediately applicable to any reparameterization of the resolution function, as the involved reconstruction procedures are entirely independent of the underlying network structure. We have applied a reconstructed resolution function to the pre-established neutron capture resonances in the $^{53}$Cr($n,\gamma$) reaction. We found an excellent agreement with the preliminary experimental data from n\_TOF, thus providing a proof of concept that the resolution function parametrization proposed here is indeed feasible.


Unlike the resolution function for the first experimental area (EAR1) of the n\_TOF facility, the one for the second experimental area (EAR2) features a strong \textit{nontrivial} dependence on the sample position, i.e. on the distance from a neutron source (a Pb spallation target). Therefore, one could parametrize several separate EAR2 resolution functions at different sample positions of interest. On the other hand, a parametrization procedure proposed here could be easily extended so as to include a sample position as an additional input parameter, alongside a true neutron energy and an effective neutron-moderation length. By extending (reoptimizing) the network structure, a comprehensive parametrization of the EAR2 resolution function -- particularly its variation along the neutron beam -- could be achieved in a single go, for a wide range of sample positions.

\section*{Acknowledgments}

This work was supported by the Croatian Science Foundation under the project number HRZZ-IP-2022-10-3878. This project has received funding from the European Union’s Horizon Europe Research and Innovation programme under Grant Agreement No 101057511.

\section*{Appendix}
\label{appendix}

We address here a visual appearance of norm violation in applying a resolution function to a unit spectrum from Fig.~\ref{fig_unit_spectrum}. Both transformed spectra -- obtained by applying either the raw or the fitted resolution function -- seem to be systematically raised above unity, implying that the total number of underlying counts might not be conserved. This is only a visual artifact caused by using a nonuniform (in particular isolethargic) binning, in combination with a logarithmic scale used for the display.

\pagebreak

We illustrate in broad stokes how the effect comes about. We denote the differential spectra of counts from Eq.~(\ref{counts_trans}) as:
\begin{equation*}
Y_\en(\en')\equiv\frac{\D N_\en(\en')}{\D\en'} \quad\text{and}\quad Y_\etof(\etof')\equiv\frac{\D N_\etof(\etof')}{\D\etof'}.
\end{equation*}
Although these spectra do not strictly correspond to a reaction yield~$Y$, using this notation will serve as a constant reminder that a reaction yield transforms in the same way, as in Eq.~(\ref{yield}).


Let us consider a contribution to a \textit{single bin} of a transformed spectrum $Y_\etof$ (bin centered at $\bar{\etof}$ and of width $\Delta\etof$) from a \textit{single bin} of an original spectrum $Y_\en$ (bin centered at $\bar{\en}$ and of width $\Delta\en$). Let $\bar{Y}_\en$ denote a discretized (histogrammed, bin-averaged) content of an original spectrum. The total number of counts within a single bin at $\bar{\en}$ equals \mbox{$\bar{Y}_\en(\bar{\en})\Delta \en$}. Of all these counts, a resolution function transfers to a target bin at $\bar{\etof}$ a relative amount of \mbox{$\bar{R}_\etof(\bar{\en},\bar{\etof})\Delta\etof$}, with $\bar{R}_\etof(\bar{\en},\bar{\etof})$ as a representative (appropriately averaged) value of a resolution function around a point \mbox{$(\bar{\en},\bar{\etof})$}. Therefore, a partial contribution to the counts ending in a target bin at $\bar{\etof}$ equals \mbox{$\bar{Y}_\en(\bar{\en})\bar{R}_\etof(\bar{\en},\bar{\etof})\Delta \en\Delta\etof$}. A partial contribution \mbox{$\mathcal{Y}_\etof(\bar{\en},\bar{\etof})$} to a discretized \textit{differential} spectrum $\bar{Y}_\etof$ needs to be normalized by $\Delta\etof$, hence: 
\begin{equation}
\mathcal{Y}_\etof(\bar{\en},\bar{\etof})=\frac{\bar{Y}_\en(\bar{\en})\bar{R}_\etof(\bar{\en},\bar{\etof})\Delta \en}{\sum_\varepsilon \bar{R}_\etof(\bar{\en},\bar{\varepsilon})\Delta \varepsilon}.
\label{y_e}
\end{equation}
The numerator represents the transfered counts normalized over a target bin width~$\Delta\etof$, while the denominator represents (in a simplified notation) a normalization of a discretized resolution function. It may at first seem superfluous, since its discretization \textit{must} be such that \mbox{$\sum_\varepsilon \bar{R}_\etof(\bar{\en},\bar{\varepsilon})\Delta \varepsilon=1$}. However, it will be a precise source of an apparent norm-violating effect from Fig.~\ref{fig_unit_spectrum}, due to a nonuniform binning resulting in a transfer of counts between the bins of unequal widths $\Delta\en$ and $\Delta\etof$.

To demonstrate this in a simplified manner, we now assume an isolethargic binning from Fig.~\ref{fig_unit_spectrum}, as it is regularly used in displaying the data over the multiple orders of magnitude. It consists of bin widths uniformly distributed over the logarithmic scale, such that\footnote{Formally, a dimensionless argument should appear under a logarithm, so we should use \mbox{$\ln(\varepsilon/\varepsilon_0)$}, with $\varepsilon_0$ as an arbitrary carrier of appropriate physical units (e.g. \mbox{$\varepsilon_0=1$ eV}). However, $\varepsilon_0$ does not affect a difference \mbox{$\Delta\ln(\varepsilon/\varepsilon_0)$}, so we use $\Delta(\ln\varepsilon)$ for simplicity.
} \mbox{$\Delta(\ln\varepsilon)=$const.} For sufficiently dense binning we may approximate a differential identity \mbox{$\D(\ln\varepsilon)=\D\varepsilon/\varepsilon$} by applying it to finite bin widths: \mbox{$\Delta(\ln\varepsilon)=\Delta\varepsilon/\bar{\varepsilon}$}. In that, we use the same binning for both types of energy, so that:
\begin{equation}
\Delta\ell=\frac{\Delta\en}{\bar{\en}}=\frac{\Delta\etof}{\bar{\etof}},
\label{d_ell}
\end{equation}
with \mbox{$\Delta\ell=\Delta(\ln\en)=\Delta(\ln\etof)$} as a \textit{constant} value characterizing a density of isolethargic binning. For further simplicity we consider a resolution function that behaves as a Dirac $\delta$-function: \mbox{$R_\etof(\en,\etof')=\delta[\etof'-\widetilde{\etof}(E)]$}. The effect of such resolution function is a complete transfer of all counts from~$\en$ to a single target value of~$\widetilde{\etof}(E)$. In a context of binned data from Eq.~(\ref{y_e}) a discretized version of such function may be obtained from a normalization condition $\bar{R}_\etof(\bar{\en},\bar{\etof}_{\bar{\en}})\Delta \etof_{\bar{\en}}=1$, with $\bar{\etof}_{\bar{\en}}$ and $\Delta \etof_{\bar{\en}}$ as a position and width of a \textit{single} bin that contains a target value~$\widetilde{\etof}(\bar{\en})$, thus exhausting an entire normalization sum. Using Eq.~(\ref{d_ell}), an isolethargically  binned differential spectrum $\bar{Y}_\etof(\bar{\etof})$ of counts transferred between bins of generally different widths \mbox{$\Delta \en=\bar{\en}\Delta\ell$} and \mbox{$\Delta \etof_{\bar{\en}}=\bar{\etof}_{\bar{\en}}\Delta\ell$} receives a following contribution from a bin at $\bar{\en}$:
\begin{equation}
\mathcal{Y}_\etof(\bar{\en},\bar{\etof}_{\bar{\en}})=\frac{\bar{R}_\etof(\bar{\en},\bar{\etof}_{\bar{\en}})\times \bar{\en}\Delta \ell}{\bar{R}_\etof(\bar{\en},\bar{\etof}_{\bar{\en}})\times \bar{\etof}_{\bar{\en}}\Delta \ell}\bar{Y}_\en(\bar{\en})=\frac{\bar{\en}}{\bar{\etof}_{\bar{\en}}}\bar{Y}_\en(\bar{\en}).
\label{amplify}
\end{equation}
Thus, a contribution to a \textit{differential isolethargic} spectrum of transferred counts is amplified by a factor $\bar{\en}/\bar{\etof}_{\bar{\en}}$ due to a difference in bin widths.

\begin{figure}[t!]
\centering
\includegraphics[width=1\linewidth]{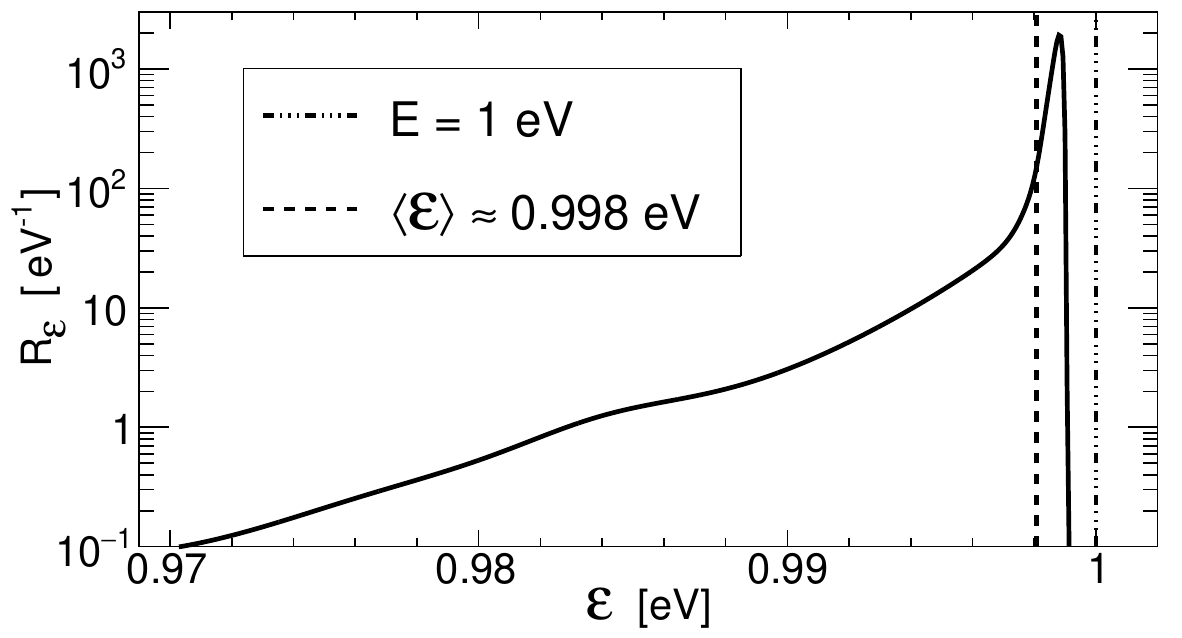}
\caption{Slice through a fitted resolution function $R_\etof(\en,\etof')$ at \mbox{$\en=1$ eV}, obtained by a transformation from Eq.~(\ref{trans_etof}).}
\label{fig_rf_etof}
\vspace{2mm}
\end{figure}


Roughly speaking -- for locally flat original spectrum~$\bar{Y}_\en$ and a resolution function slowly varying with~$\en$ -- one can expect a total spectrum~$\bar{Y}_\etof$ of transferred counts to be amplified by an average ratio \mbox{$\langle \en/\etof\rangle$}, averaged over all neutron energies~$\en$ contributing to a transferred content at~$\etof$. In turn -- with all these assumptions satisfied -- this ratio may be estimated by observing an average value $\langle\etof\rangle$ for a single value of $\en$ (i.e. by observing a resolution function slice at a given value of~$E$). This allows us to approximate: 
\begin{equation}
\langle \en/\etof\rangle_\text{at $\etof$,over $\en$}\approx E/\langle\etof\rangle_\text{at $\en$, over $\etof$}.
\end{equation}
As a verification of this procedure, Fig.~\ref{fig_rf_etof} shows a resolution function slice at \mbox{$\en=1$ eV}, obtained by a transformation from Eq.~(\ref{trans_etof}). An average value of reconstructed neutron energies is \mbox{$\langle\etof\rangle\approx0.998$ eV}. Hence, around 1~eV one might expect an amplification of a transformed spectrum by a factor of 1.002, which is perfectly consistent with Fig.~\ref{fig_unit_spectrum}.


\begin{figure}[t!]
\centering
\begin{overpic}[width=1\linewidth]{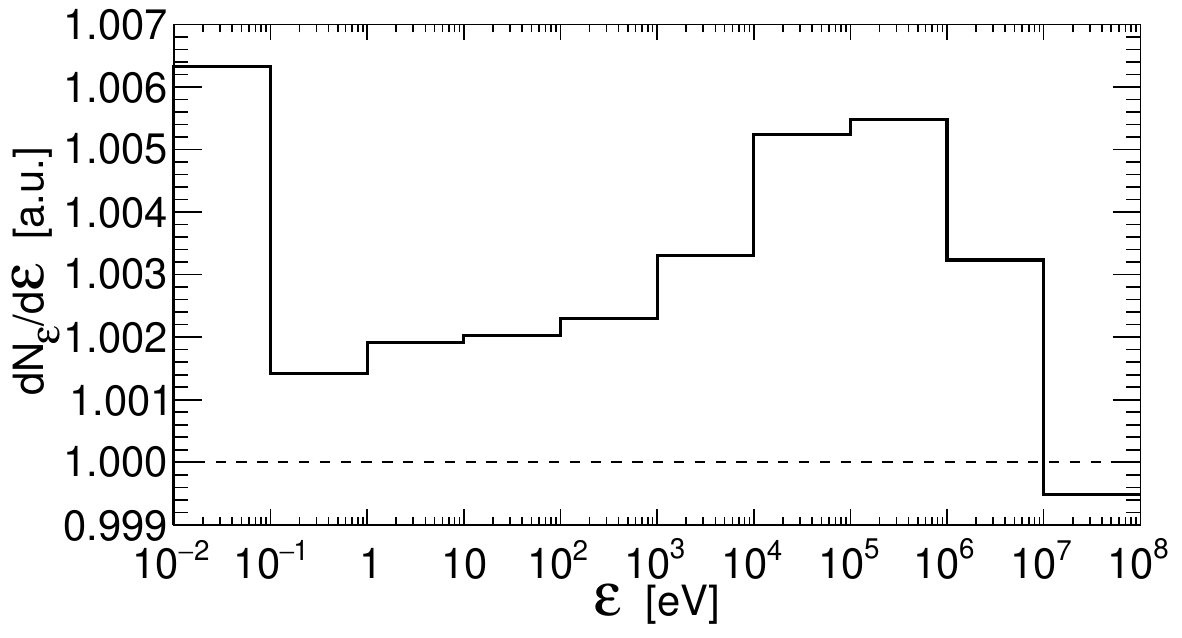}
\put(24,33){\includegraphics[width=0.39\linewidth]{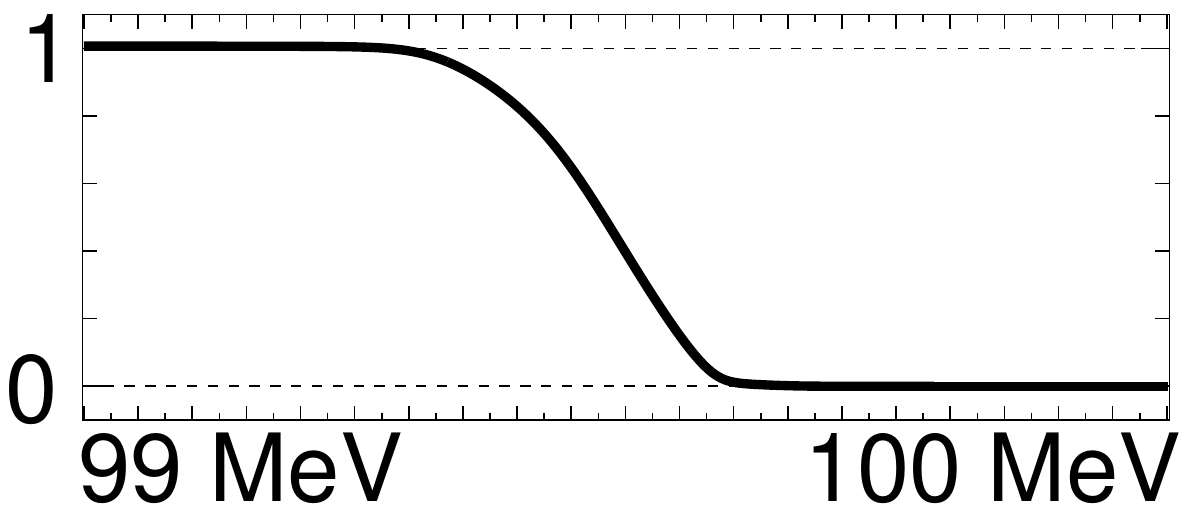}}
\end{overpic}
\caption{Resolution function transformed spectrum from Fig.~\ref{fig_unit_spectrum} in a coarse binning of 1 bin per decade (the two spectra from Fig.~\ref{fig_unit_spectrum} are now indistinguishable). Inset shows, in very fine binning, a closeup of a high end spectrum between 99~MeV and 100~MeV, perfectly compensating for a slight systematic increase throughout the lower part of the spectrum, thus justifying a norm preservation.}
\label{fig_unit_coarse}
\end{figure}

We have justified a seemingly-systematic amplification of a transformed differential spectrum due to a use of isolethargic binning. This does not yet resolve the issue of a seeming norm violation, i.e. of an apparent nonconservation of counts from Fig.~\ref{fig_unit_spectrum}. \textit{The norm is indeed preserved.} However, it does not appear so due to a use of a logarithmic scale in displaying the data. An amplification from Eq.~(\ref{amplify}) \textit{for those bins whose content is amplified} may be summarized like this. As illustrated in Fig.~\ref{fig_rf_etof}, a resolution function transfers the counts from a higher value of~$\en$ to the lower values of~$\etof$ (at least on average). Assuming approximately flat and isolethargically binned original spectrum~$Y_\en$, this means that the counts from wider $\Delta\en$ bins are, on average, transfered to narrower $\Delta \etof$ bins. Renormalizing the transferred counts due to an average decrease in bin width leads to an amplification of a differential spectrum. In that, most energy-bins lose some counts by their transfer to lower $\etof$~vales, while obtaining some counts from higher $\en$~values, balancing a final amplification. However, this balanced loss-gain interplay no longer applies to the high end of the spectrum, where the counts are mostly lost due to no counts existing at higher energies. Therefore, one should expect a slight increase in the greater part of the spectrum to be compensated by a disproportionate decrease at the high end of it. This is indeed the case. It is not apparent in Fig.~\ref{fig_unit_spectrum} because the high end of the spectrum is \textit{visually suppressed} in logarithmic scale, while the fine binning constrains the effect to the last few bins. To prove this claim, Fig.~\ref{fig_unit_coarse} shows the same (appropriately renormalized) spectrum in a coarse binning of 1 bin per decade. A single spectrum is shown due to two spectra from Fig~\ref{fig_unit_spectrum} (obtained from the raw or the fitted resolution function) being indistinguishable in such coarse binning. An expected decrease in the highest energy bin is clearly visible. It is perfectly sufficient to compensate an increase in the rest of the spectrum, since the last bin (from 10~MeV to 100~MeV) is 9~times wider than an entire preceding part of the spectrum (below 10~MeV). As the source of this decrease is hardly visible in Fig~\ref{fig_unit_spectrum}, an inset shows a very densely binned closeup at the extreme end of the spectrum, between 99~MeV and 100~MeV. \pagebreak

\noindent A decrease from approximate unity starts around 99.3~MeV and at 99.6~MeV reaches approximate zero. This drastic but narrow drop is altogether sufficient to compensate for a slight systematic increase across the entire lower part of the spectrum. We remind the reader that such narrow decrease is obtained by using a proton-beam-width-free form of a resolution function, as in Fig.~\ref{fig_unit_spectrum}. When using a realistic value of \mbox{$\sigma_T=7$~ns}, a decrease region is smeared from approximately 97~MeV to 102~MeV.\linebreak

In conclusion, a visual estimation of an area below a spectrum displayed in logarithmic scale may be misleading in appraising the true area (i.e. its norm), because such visual area is given by the integral of the form \mbox{$\int Y_\varepsilon(\varepsilon') \D(\log_{10}\varepsilon')\propto \int[\varepsilon' Y_\varepsilon(\varepsilon')]\D\varepsilon'$}. Evidently, the integrand \mbox{$\varepsilon' Y_\varepsilon(\varepsilon')$} is very much affected relative to $Y_\varepsilon(\varepsilon')$ from a true area expression  \mbox{$\int Y_\varepsilon(\varepsilon')\D\varepsilon'$}, thus necessitating extreme care in drawing certain types of conclusions from a logarithmic plot.

{\color{white}
.

.
}





\begin{thebibliography}{9}

\bibitem{conception} C. Rubbia, S. Andriamonje, D. Bouvet-Bensimon et al., A high resolution spallation driven facility at the CERN-PS to measure neutron cross sections in the interval from 1 eV to 250 MeV. (CERN/LHC/98-02 and CERN/LHC/98-02-Add.1, 1998), \href{https://cds.cern.ch/record/363828}{https://cds.cern.ch/record/363828}

\bibitem{facility} C. Guerrero, A. Tsinganis, E. Berthoumieux et al., Performance of the neutron time-of-flight facility n\_TOF at CERN. Eur. Phys. J. A \textbf{49}, 27 (2013). \href{https://doi.org/10.1140/epja/i2013-13027-6}{{\DOI}10.1140/epja/i2013-13027-6}

\bibitem{phase4} N. Patronis, A. Mengoni, S. Goula et al., Status report of the n\_TOF facility after the 2nd CERN long shutdown period. EPJ Tech. Instrum. \textbf{10}, 13 (2023). \href{https://doi.org/10.1140/epjti/s40485-023-00100-w}{{\DOI}10.1140/epjti/s40485-023-00100-w}
\bibitem{phase4_target} R. Esposito, M. Calviani, O. Aberle et al., Design of the third-generation lead-based neutron spallation target for the neutron time-of-flight facility at CERN. Phys. Rev. Accel. Beams \textbf{24}, 093001 (2021). \href{https://doi.org/10.1103/PhysRevAccelBeams.24.093001}{{\DOI}10.1103/PhysRevAccelBeams.24.093001}

\bibitem{ear2_1} C. Wei\ss , E. Chiaveri, S. Girod et al., The new vertical neutron beam line at the CERN n\_TOF facility design and outlook on the performance. Nucl. Instrum. Methods Phys. Res. A \textbf{799}, 90--98 (2015). \href{https://doi.org/10.1016/j.nima.2015.07.027}{{\DOI}10.1016/j.nima.2015.07.027}
\bibitem{ear2_2} N. Colonna, E. Chiaveri, F. Gunsing, The Second Beam-Line and Experimental Area at n\_TOF: A New Opportunity for Challenging Neutron Measurements at CERN. Nucl. Phys. News \textbf{25}(4) 19--23 (2015). \href{https://doi.org/10.1080/10619127.2015.1035930}{{\DOI}10.1080/10619127.2015.1035930}
\bibitem{ear2_3} S. Barros, I. Bergstr\"{o}m, V. Vlachoudis, C. Wei\ss, Optimization of n\_TOF-EAR2 using FLUKA. J. Instrum. \textbf{10}, P09003 (2015). \href{https://doi.org/10.1088/1748-0221/10/09/P09003}{{\DOI}10.1088/1748-0221/10/09/P09003}

\bibitem{near_1} M. Ferrari, D. Senajova, O. Aberle et al., Design development and implementation of an irradiation station at the neutron time-of-flight facility at CERN. Phys. Rev. Accel. Beams \textbf{25}, 103001 (2022). \href{https://doi.org/10.1103/PhysRevAccelBeams.25.103001}{{\DOI}10.1103/PhysRevAccelBeams.25.103001}
\bibitem{near_2} N. Patronis, A. Mengoni, N. Colonna et al., The CERN n\_TOF NEAR station for astrophysics- and application-related neutron activation measurements. (arXiv:2209.04443 [physics.ins-det], 2022), \href{https://arxiv.org/abs/2209.04443}{https://arxiv.org/abs/2209.04443}
\bibitem{near_3} M. E. Stamati, P. Torres-S\'{a}nchez, P. P\'{e}rez-Maroto et al., The n\_TOF NEAR Station Commissioning and first physics case. EPJ Web Conf. \textbf{284}, 06009 (2023). \href{https://doi.org/10.1051/epjconf/202328406009}{{\DOI}10.1051/epjconf/202328406009}

\pagebreak

\bibitem{flux_ear1} M. Barbagallo, C. Guerrero, A. Tsinganis et al., High-accuracy determination of the neutron flux at n\_TOF. Eur. Phys. J. A \textbf{49}, 156 (2013). \href{https://doi.org/10.1140/epja/i2013-13156-x}{{\DOI}10.1140/epja/i2013-13156-x}
\bibitem{flux_ear2} M.~Sabat\'{e}-Gilarte, M. Barbagallo, N. Colonna et al., High-accuracy determination of the neutron flux in the new experimental area n\_TOF-EAR2 at CERN. Eur. Phys. J. A \textbf{53}, 210 (2017). \href{https://doi.org/10.1140/epja/i2017-12392-4}{{\DOI}10.1140/epja/i2017-12392-4}

\bibitem{beam1} S. Lo Meo, M. A. Cort\'{e}s-Giraldo, C. Massimi et al., GEANT4 simulations of the n\_TOF spallation source and their benchmarking. Eur. Phys. J. A \textbf{51}, 160 (2015). \href{https://doi.org/10.1140/epja/i2015-15160-6}{{\DOI}10.1140/epja/i2015-15160-6}
\bibitem{beam2} J. Lerendegui-Marco, S. Lo Meo, C. Guerrero et al., Geant4 simulation of the n\_TOF-EAR2 neutron beam: Characteristics and prospects. Eur. Phys. J. A \textbf{52}, 100 (2016). \href{https://doi.org/10.1140/epja/i2016-16100-8}{{\DOI}10.1140/epja/i2016-16100-8}

\bibitem{start_1} C. Coceva, M. Frisoni, M. Magnani, A. Mengoni, On the figure of merit in neutron time-of-flight measurements. Nucl. Instrum. Methods Phys. Res. A \textbf{489}, 346--356 (2002). \href{https://doi.org/10.1016/S0168-9002(02)00903-8}{{\DOI}10.1016/S0168-9002(02)00903-8}
\bibitem{start_2} C. Borcea, P. Cennini, M. Dahlfors et al., Results from the commissioning of the n\_TOF spallation neutron source at CERN. Nucl. Instrum. Methods Phys. Res. A \textbf{513}, 524--537 (2003). \href{https://doi.org/10.1016/S0168-9002(03)02072-2}{{\DOI}10.1016/S0168-9002(03)02072-2}

\bibitem{rf_ntof} V. Vlachoudis, M. Sabate-Gilarte, V. Alcayne et al., On the resolution function of the n\_TOF facility: a comprehensive study and user guide. (n\_TOF-PUB-2021-001, 2021), \href{https://cds.cern.ch/record/2764434/}{https://cds.cern.ch/record/2764434}

\bibitem{root} Rene Brun, Fons Rademakers, ROOT -- An object oriented data analysis framework. Nucl. Instrum. Methods Phys. Res. A \textbf{389}, 81--86 (1997). \href{https://doi.org/10.1016/S0168-9002(97)00048-X}{{\DOI}10.1016/S0168-9002(97)00048-X}

\bibitem{phases_1} C. Domingo-Pardo, O. Aberle, V. Alcayne et al., The neutron time-of-flight facility n\_TOF at CERN Recent facility upgrades and detector developments. J. Phys. Conf. Ser. \textbf{2586}, 012150 (2023). \href{https://doi.org/10.1088/1742-6596/2586/1/012150}{{\DOI}10.1088/1742-6596/2586/1/012150}

\bibitem{phases_2} E. Dupont, N. Otuka, D. Rochman et al., Overview of the dissemination of n\_TOF experimental data and resonance parameters. EPJ Web Conf. \textbf{284}, 18001 (2023). \href{https://doi.org/10.1051/epjconf/202328418001}{{\DOI}10.1051/epjconf/202328418001}

\bibitem{rf_unfolding}  P. \v{Z}ugec, N. Colonna, M. Sabate-Gilarte et al., A direct method for unfolding the resolution function from measurements of neutron induced reactions. Nucl. Instrum. Methods Phys. Res. A \textbf{875}, 41--50 (2017). \href{https://doi.org/10.1016/j.nima.2017.09.004}{{\DOI}10.1016/j.nima.2017.09.004} \pagebreak


\bibitem{approximators1} K. Hornik, M. Stinchcombe, H. White, Multilayer feedforward networks are universal approximators. Neural Netw. \textbf{2}, 359--366 (1989). \href{https://doi.org/10.1016/0893-6080(89)90020-8}{{\DOI}10.1016/0893-6080(89)90020-8}

\bibitem{approximators2} M. Leshno, V. Ya. Lin, A. Pinkus, S. Schocken, Multilayer feedforward networks with a nonpolynomial activation function can approximate any function. Neural Netw. \textbf{6}, 861--867 (1993). \href{https://doi.org/10.1016/S0893-6080(05)80131-5}{{\DOI}10.1016/S0893-6080(05)80131-5}

\bibitem{neural_2021} V. Babiano-Su\'{a}rez, J. Lerendegui-Marco, J. Balibrea-Correa et al., Imaging neutron capture cross sections: i-TED proof-of-concept and future prospects based on Machine-Learning techniques. Eur. Phys. J. A \textbf{57}, 197 (2021). \mbox{\href{https://doi.org/10.1140/epja/s10050-021-00507-7}{{\DOI}10.1140/epja/s10050-021-00507-7}}

\bibitem{neural_2022} P. \v{Z}ugec, M. Barbagallo, J. Andrzejewski et al., Machine learning based event classification for the energy-differential measurement of the $^\text{nat}$C(\textit{n,p}) and $^\text{nat}$C(\textit{n,d}) reactions. Nucl. Instrum. Methods Phys. Res. A \textbf{1033}, 166686 (2022). \href{https://doi.org/10.1016/j.nima.2022.166686}{{\DOI}10.1016/j.nima.2022.166686}

\bibitem{neural_2023} A. Sanchez-Caballero, V. Alcayne, D. Cano-Ott et al., A Case Study on Deep Learning applied to Capture Cross Section Data Analysis. EPJ Web Conf. \textbf{284}, 16001 (2023). \href{https://doi.org/10.1051/epjconf/202328416001}{{\DOI}10.1051/epjconf/202328416001}

\bibitem{rf_neural} T. Cavagna, n\_TOF Transport Code Update and RF Deconvolution. (CERN-STUDENTS-Note-2023-101, 2023), \href{https://cds.cern.ch/record/2869067/}{https://cds.cern.ch/record/2869067}

\bibitem{root_perceptron} CERN ROOT: TMultiLayerPerceptron Class, \href{https://root.cern/doc/master/classTMultiLayerPerceptron.html}{https:// root.cern/doc/master/classTMultiLayerPerceptron.html}. Accessed 14 March 2025

\bibitem{data_bank} P. \v{Z}ugec, M. Sabate Gilarte, M. Bacak et al., Resolution function data for n\_TOF EAR1 Phase-3 [DS/OL]. V2. Science Data Bank, 2025 [2025-05-23]. \href{https://doi.org/10.57760/sciencedb.j00186.00697}{{\DOI}10.57760/sciencedb.j00186.00697}. Accessed 23 May 2025  \\

\bibitem{cr53} P. P\'{e}rez-Maroto, C. Guerrero, A. Casanovas et al., Description and outlook of the $^{50,53}$Cr($n,\gamma$) cross section measurement at n\_TOF and HiSPANoS. EPJ Web Conf. \textbf{294}, 01004 (2024). \href{https://doi.org/10.1051/epjconf/202429401004}{{\DOI}10.1051/epjconf/202429401004}

\bibitem{endf8} D. A. Brown, M. B. Chadwick, R. Capote et al., ENDF/B-VIII.0: The 8$^\text{th}$ Major Release of the Nuclear Reaction Data Library with CIELO-project Cross Sections, New Standards and Thermal Scattering Data. Nucl. Data Sheets \textbf{148}, 1--142 (2018). \href{https://doi.org/10.1016/j.nds.2018.02.001}{{\DOI}10.1016/j.nds.2018.02.001}

\bibitem{bpd_5000_1} K. Fraval, F. Gunsing, S. Altstadt et al., Measurement and analysis of the $^{241}$Am($n,\gamma$) cross section with liquid scintillator detectors using time-of-flight spectroscopy at the n\_TOF facility at CERN. Phys. Rev. C \textbf{89}, 044609 (2014). \href{https://doi.org/10.1103/PhysRevC.89.044609}{{\DOI}10.1103/PhysRevC.89.044609}
\bibitem{bpd_5000_2} F. Mingrone, C. Massimi, G. Vannini et al., Neutron capture cross section measurement of $^{238}$U at the CERN n\_TOF facility in the energy region from 1 eV to 700 keV. Phys. Rev. C \textbf{95}, 034604 (2017). \href{https://doi.org/10.1103/PhysRevC.95.034604}{{\DOI}10.1103/PhysRevC.95.034604}
\bibitem{bpd_2000} V. Michalopoulou, A. Stamatopoulos, M. Diakaki et al., Measurement of the neutron-induced fission cross section of $^{230}$Th at the CERN n\_TOF facility. Phys. Rev. C \textbf{108}, 014616 (2023). \href{https://doi.org/10.1103/PhysRevC.108.014616}{{\DOI}10.1103/PhysRevC.108.014616}

\bibitem{rf_guide} P. \v{Z}ugec, User guide through Resolution Function class. (n\_TOF-PUB-2025-001, 2025), \href{http://cds.cern.ch/record/2926718/}{http://cds.cern.ch/record/2926718} 

\bibitem{fft} J. W. Cooley, J. W. Tukey, An algorithm for the machine calculation of complex Fourier series. Math. Comput. \textbf{19}, 297--301 (1965). \href{https://doi.org/10.1090/S0025-5718-1965-0178586-1}{{\DOI}10.1090/S0025-5718-1965-0178586-1}

\bibitem{numc} W. H. Press, S. A. Teukolsky, W. T. Vetterling, B. P. Flannery, \textit{Numerical Recipes in C: The Art of Scientific Computing}, 3rd edn. (Cambridge University Press, Cambridge New York Melbourne Madrid Cape Town Singapore S\~{a}o Paulo, 2007), pp. 132--133, 608--620


\end{thebibliography}
\end{document}